%% file: Les.proc.6XI14.tex
\documentclass[12pt]{article}
\usepackage{graphicx}

\def\Title#1{\begin{center} {\Large #1 } \end{center}}
\def\Author#1{\begin{center}{ \sc #1} \end{center}}
\def\Address#1{\begin{center}{ \it #1} \end{center}}

\newenvironment{Abstract}{\begin{quotation}  }{\end{quotation}}
\newenvironment{Presented}{\begin{quotation} \begin{center} 
             PRESENTED AT\end{center}\bigskip 
      \begin{center}\begin{large}}{\end{large}\end{center} \end{quotation}}
\def\Acknowledgements{\bigskip  \bigskip \begin{center} \begin{large}
             \bf ACKNOWLEDGEMENTS \end{large}\end{center}}

\input econfmacros.tex

\begin{document}
\begin{titlepage}

\vfill
\Title{Dalitz plot studies in hadronic charm decays}
\vfill
\Author{Leonard Le\'sniak
}
\Address{Division of Theoretical Physics, The Henryk Niewodnicza\'nski Institute of Nuclear Physics,
                  Polish Academy of Sciences, 31-342 Krak\'ow, POLAND\\}
\vfill
\begin{Abstract}
Recent studies of hadronic $D$-meson decays are reported.
Some experimental searches of $CP$-symmetry violation using model independent
methods are presented. 
An importance of unitarity constraints in construction of phenomenological models
of the $D$-meson decays is underlined.
The theoretical model of the $D^0 \to K^0_S \pi^+ \pi^-$ decays, including some two-body unitarity 
constraints, is described. Then a comparison of the model results with the Belle collaboration data
is made. The results on the $CP$-violation in the $D^0 \to K^0_S \pi^+ \pi^-$ decays are given
and the necessity to consider the $CP$-violation in the subsequent $K^0_S$ decays is emphasized. 
 
\end{Abstract}
\vfill
\begin{Presented}
the 8th International Workshop
on the CKM Unitarity Triangle (CKM 2014), Vienna, Austria, September 8-12, 2014
\end{Presented}
\vfill
\end{titlepage}
\def\thefootnote{\fnsymbol{footnote}}
\setcounter{footnote}{0}

\section{Introduction}
Studies of the charm meson decays into multimeson final states have many aspects. 
In particular, one can indicate measurements of the $D^0- \bar D^0$ mixing parameters
and the Cabibbo-Kobayashi-Maskawa angle $\gamma$ (or $\phi_3$), and searches of $CP$ 
violation. Understanding the final state strong interactions between the produced particles,
development of the model independent methods or a construction of the models satisfying unitarity, 
analyticity and chiral symmetry constraints are other important topics. Also improving the
isobar models frequently used in analyses of experimental data is highly desirable.

\section{Search for \mbox{\boldmath $CP$} violation in model independent Dalitz plot analyses}
Density distributions of events in the Dalitz plots are studied in three-body $D$-meson decays.
Recently, binned or unbinned methods in searches for $CP$ violation effects in the model independent Dalitz plot 
analyses are applied.
Different variables are used in these studies.
 For example, in the BABAR collaboration analysis of the $D^{\pm}\to K^+K^-\pi^{\pm}$
decays~\cite{lees} the so-called normalized residual variables $\Delta_i$ are used
\begin{equation}
\Delta_i = \frac{n_i(D^+)-n_i(D^-)}{\sqrt{\sigma_i^2(D^+)+R^2\sigma_i^2(D^-)}}, 
\end{equation}
where $n_i(D^+),n_i(D^-)$ are numbers of signal events in the $i$-th bin of the Dalitz plot,
$\sigma_i^2(D^+)$ and $\sigma_i^2(D^-)$ are the statistical uncertainties.
The correction factor, $R=\sum n_i(D^+)/ \sum n_i(D^-)$, is used to remove the production or detection
asymmetries.
In the recent publication~\cite{LHCb} devoted to the search of the $CP$ violation 
in the $D^{\pm}\to \pi^+\pi^-\pi^{\pm}$ 
decays the LHCb collaboration has used a similar variable $S^i_{CP}$, called significance:
\begin{equation}
S^i_{CP} = \frac{n_i(D^+)-n_i(D^-)}{\sqrt{R[n_i(D^+)+n_i(D^-)]}}.
\end{equation}
In absence of $CP$ violation the distributions of the variables $\Delta_i$ and $S^i_{CP}$ are 
standard normal Gaussian functions. 
In Ref.~\cite{LHCb} the LHCb collaboration has also used the unbinned $k$-nearest neighbour method.
In this technique one chooses $n_k$ nearest neighbour events in the combined $D^+$ and $D^-$ samples
with $N^+$ and $N^-$ events. Then the test variable $T$ is defined as
\begin{equation}
T=\frac{1}{n_k(N^+ + N^-)} \sum_{i=1}^{N^+ + N^-} \sum_{k=1}^{n_k} I(i,k),
\end{equation}
where $I(i,k)=1$ if the $i$-th event and its $k$-th nearest neighbour have the same $D$ charge
and $I(i,k)=0$ otherwise. In absence of $CP$ asymmetry the $T$-distribution is Gaussian with known
mean and variance values. 

Using both binned and unbinned methods the LHCb collaboration has
found no evidence for $CP$ violation in the $D^{\pm}\to \pi^+\pi^-\pi^{\pm}$ 
decays~\cite{LHCb}. Also the BABAR collaboration result for the $D^{\pm}\to K^+K^-\pi^{\pm}$
decays was negative~\cite{lees}. Let us remark here that if the $CP$ violation is not found then 
there is no model-independent way to find its upper limit. On the other hand, upper limits can be 
determined in analyses in which a model of decay amplitudes is constructed. 


\section{Model of the \mbox{\boldmath $D^0 \to K^0_S \pi^+ \pi^-$} decays with two-body unitarity
 constraints}
The so-called isobar model is widely used in many experimental analyses of the $D$-meson decays. 
However, it is not 
unitary. Therefore determinations of the branching fractions and the $CP$ asymmetries could not be
sufficiently accurate if such a model is applied. In order to improve a description of data  
an attempt to incorporate two-body unitarity constraints 
in the model of the $D^0 \to K^0_S \pi^+ \pi^-$ decay amplitudes has been done for 
the following subchannels: $K^0_S \pi$ $S$-wave, 
$\pi \pi$ $S$-wave and $\pi \pi$ $P$-wave~\cite{DKLL}. 

\begin{figure}[htb]
\centering
\includegraphics[scale = 0.39]{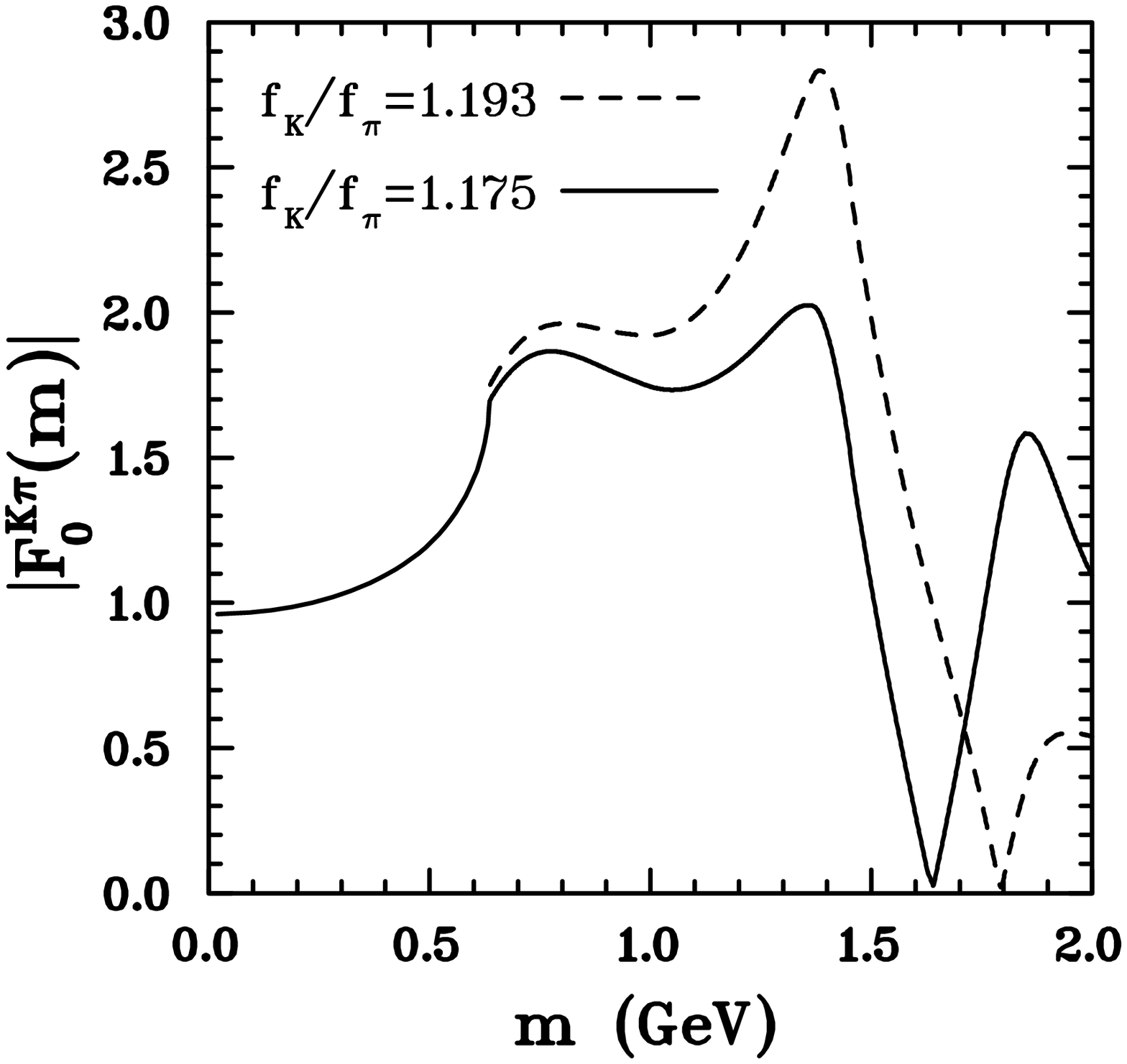}~~~~
\includegraphics[scale = 0.39]{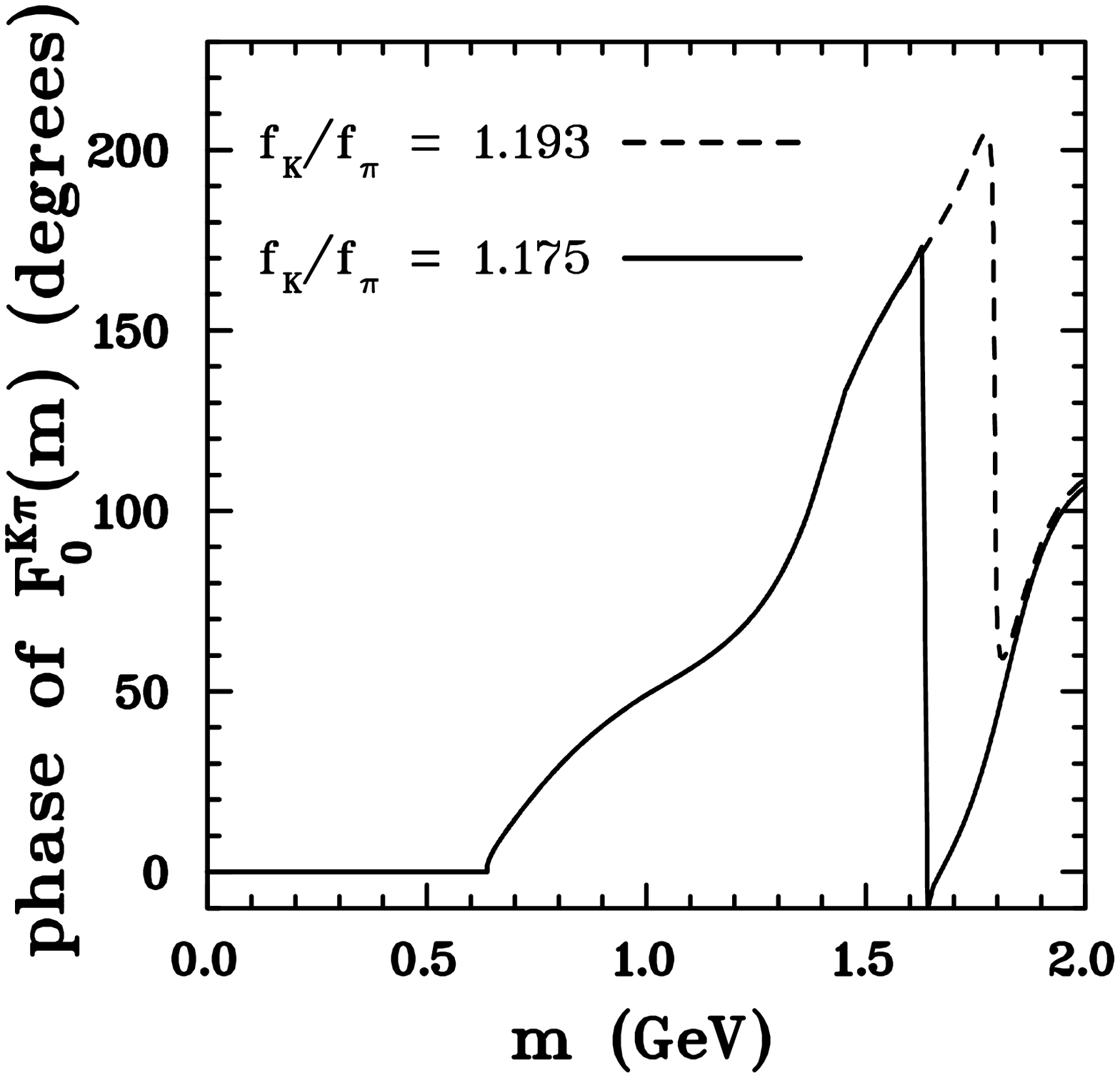}~~~~
\caption{The modulus (left panel) and the phase (right panel) of the $K\pi$ scalar form factor
as function of the $K\pi$ effective mass $m$ for two values of the $f_K/f_{\pi}$ ratio where $f_K$
and $f_{\pi}$ are the kaon and the pion decay constants.}
\label{fig:Kpi}
\end{figure}

A factorization approximation has been used in calculations of the decay 
amplitudes. 
The annihilation (via $W$-exchange) amplitudes have been added to the tree weak decay amplitudes. 
Strong
interactions between the kaon-pion and pion-pion pairs in the $S$- and $P$-wave states have been 
described in terms of the corresponding form factors.
The kaon-pion and pion-pion scalar form factors have been constrained using unitarity, analyticity,
chiral symmetry and the data on the meson-meson scattering coming from other studies than those on 
weak decays of heavy mesons. 
Using these constraints twenty seven nonzero amplitudes have been combined into ten effective 
independent amplitudes. Some of these amplitudes group several meson-meson resonances. For example,
in the pion scalar form factor the three scalar resonances $f_0(500)$, $f_0(980)$ and $f_0(1400)$
are present. Similarly, in the $K\pi$ scalar form factor two strange scalar resonances $K_0^*(800)$
and $K_0^*(1430)$ contribute. This is illustrated in Figure~\ref{fig:Kpi}. The first maximum
of the function seen in the left panel corresponds to the $K_0^*(800)$ resonance and the second one
to the $K_0^*(1430)$. Both are of comparable heights.

In the $P$-wave amplitudes the strange resonances $K^*(892)$ and $K^*(1410)$ are included
as well as $\rho(770)$, $\rho(1450)$ and $\omega(782)$. 
The tensor resonances $K^*_2(1430)$ and $f_2(1270)$ are also present and the corresponding amplitudes
 are parameterized by the relativistic Breit-Wigner functions.
The thirty three free model parameters are fitted to the Belle collaboration data for the
$D^0 \to K^0_S \pi^+ \pi^-$ decays~\cite{Poluektow}. In order to constrain 
consistently the $K^*(892)$ mass and width the Belle data on the $\tau^- \to K^0_S \pi^-\nu_{\tau}$ 
decays~\cite{Epifanow} have been simultaneously fitted. Also the experimental total branching fraction
$Br(D^0 \to K^0_S \pi^+ \pi^-)=(2.82\pm0.19)$ $\%$ has been included in the $\chi^2$ fit.
Let us remind here that in a typical isobar-model application the absolute normalization of amplitudes
 is arbitrary.
 The number of degrees of freedom
in the fit was $ndf=6321+89+1-33=6378$. The first number, 6321, is equal to the total number of the 
effective cells covering the Dalitz plot of the $D^0 \to K^0_S \pi^+ \pi^-$ reaction.
 The second number, 89, corresponds to the number of bins of the $K^0_S \pi^-$
effective mass distribution measured in the $\tau^- \to K^0_S \pi^-\nu_{\tau}$ decays. The result of 
the best fit is $\chi^2=9451$ which gives $\chi^2/ndf=1.48$.
 
\begin{figure}[htb]
\centering
  \includegraphics[height=.28\textheight]{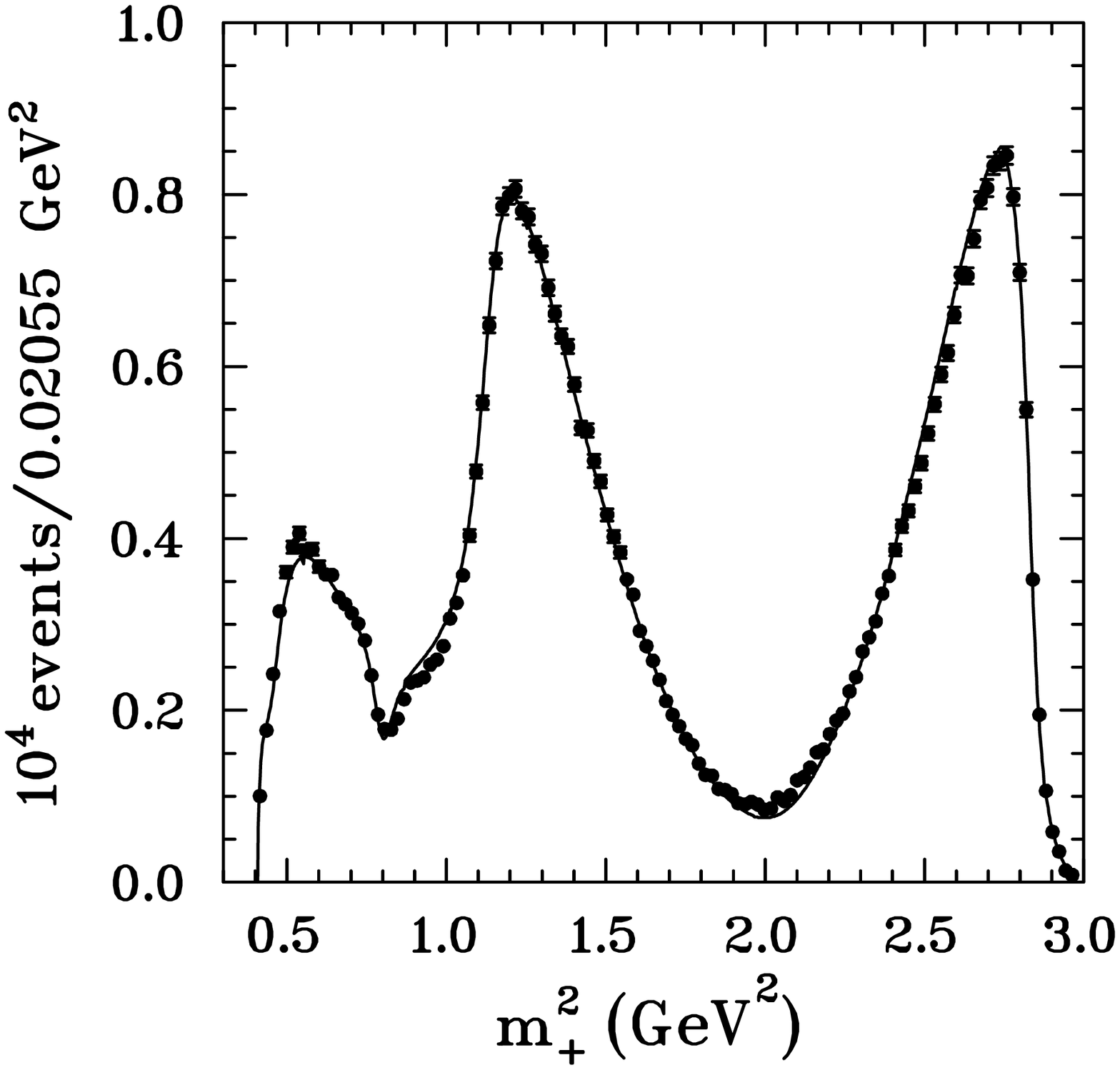}~~~~
   \includegraphics[height=.28\textheight]{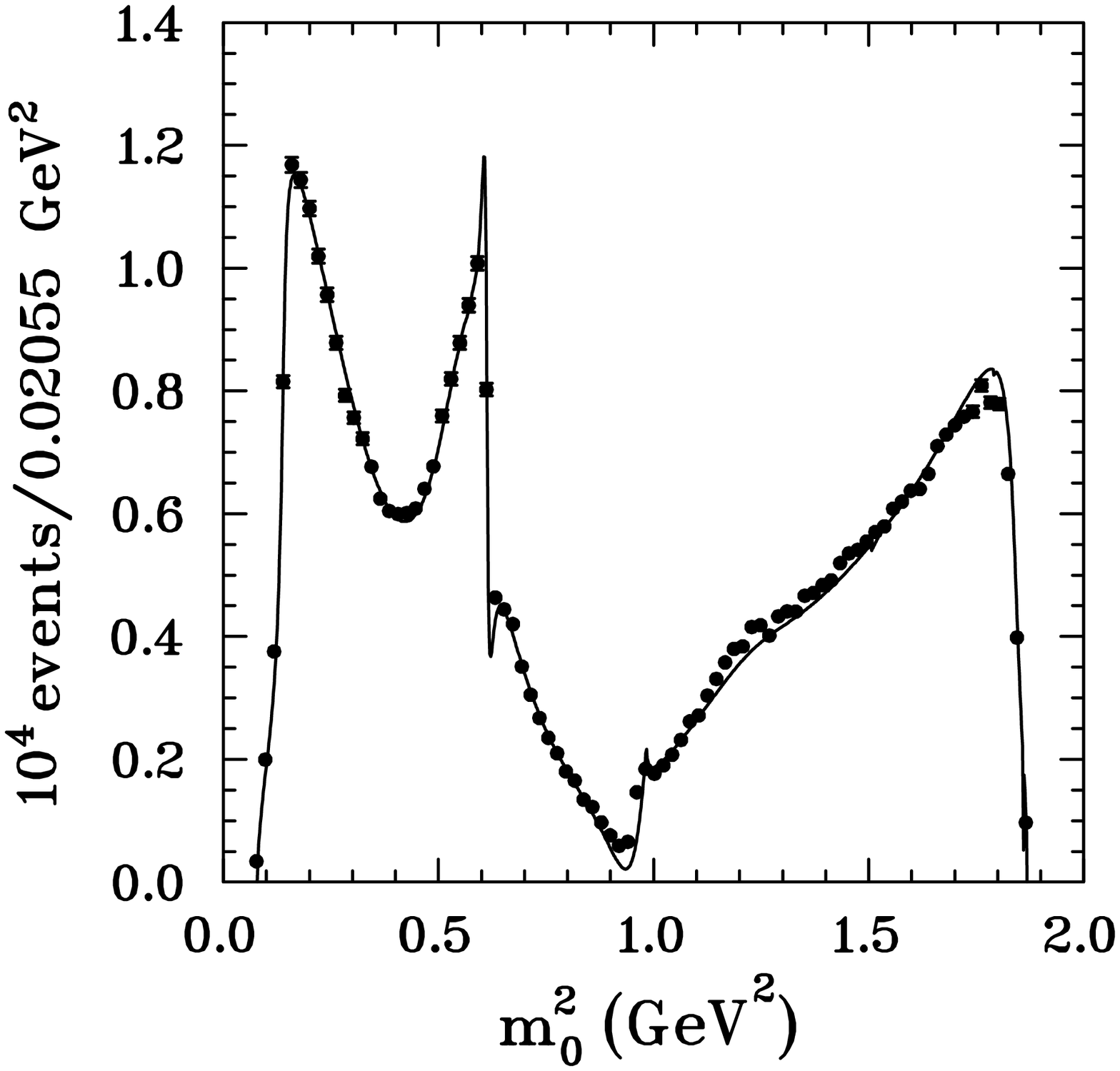}~~
\caption{Left panel: comparison of the model $K^0_S\pi^+$ effective mass squared distributions
(solid curve) with the Belle data~\cite{Poluektow}. 
Right panel: as in the left panel but for the $\pi^+\pi^-$ effective mass squared $m_0^2$.}
\label{fig:distr}
\end{figure}

A quality of the fit is shown in Figure~\ref{fig:distr} where two projections of the experimental Dalitz 
density distributions are plotted together with theoretical curves. 

The resulting parameters and tables
of branching fractions for different quasi-two-body channels can be found in Ref.~\cite{DKLL}.
An important result of the fit is a determination of the lowest values of the  
branching fractions corresponding to the annihilation amplitudes. These annihilation contributions were found 
to be relatively large.  
For example, for the $K^0_S\pi^-$ $S$-wave decay amplitude the tree branching fraction is equal to 
$8.2$ $\%$ while for the lowest value of the annihilation branching fraction one gets $7.8$ $\%$. 
Both tree and annihilation amplitudes interfere and the resulting branching fraction for this particular 
channel is equal to $25.0$ $\%$.

The theoretical model described above has also been compared with the results of the BABAR 
collaboration for the $D^0 \to K^0_S \pi^+ \pi^-$ decays~\cite{BABAR}. This time the model has
been fitted to the Dalitz plot density 
distribution obtained from the isobar model employed by BABAR to parametrize the $D^0$ decay 
amplitude.
The resulting parameters were consistent with those of the previous fit to the Belle data. 

\section{$CP$ violation in \mbox{\boldmath $D^0 \to K^0_S \pi^+ \pi^-$} decays}
The CDF collaboration has measured the time-integrated $CP$-violation asymmetry $A_{CP}$
in the $D^0 \to K^0_S \pi^+ \pi^-$ decays~\cite{CDF}. 
Using the isobar model to describe the Dalitz plot
density distribution the following result has been obtained:  $A_{CP}=(-0.05\pm0.57\pm0.54)~\%$, where
the first error is statistical and the second one is systematic. 
This result is consistent with no
$CP$-violation effect and it is consistent with the standard model prediction. 
The CDF collaboration has also performed a model-independent bin-by-bin comparison of the $D^0$ and
$\bar D^0$ Dalitz plots and again no significant $CP$ violation has been seen. 

The Belle collaboration has recently measured the $D^0-\bar D^0$ mixing parameters and searched for 
the indirect 
$CP$ symmetry violation in mixing and in the interference between mixing and decay~\cite{Peng}. 
The results are again consistent with no $CP$ violation.

The $CP$-asymmetry in the theoretical model described in the previous section  is very small, of the 
order of $5\cdot10^{-3}$~$\%$. 
It is obtained if the $CP$-violation in the $K^0_S$ decays  is neglected.
The above low value is related to the small imaginary part of the 
Cabibbo-Kobayashi-Maskawa quark matrix element $V_{cd}$ which in the standard model is proportional  
to $\lambda^5$, $\lambda$ being the Wolfenstein parameter equal to about 0.225.
However, if one takes into account the $CP$-violation in the $K^0_S$ decays then the full 
$CP$-asymmetry
can be estimated to be of the order of $-0.3\%$. This value is comparable to the present 
experimental uncertainties of the $CP$ violation measurements. Thus the $CP$ violation effects
in $K^0_S$ decays have to be taken into account in searches for $CP$ asymmetry in the 
$D^0 \to K^0_S \pi^+ \pi^-$ decays. 

\section{Conclusions}

$CP$ violation in $D$-meson decays has not yet been discovered. In future searches
both model-independent and model-dependent methods should be used to study
the Dalitz plot distributions of the hadronic charm meson decays. The model-independent methods
can be very effective in finding the local $CP$-asymmetries in different regions of the Dalitz plots.
On the other hand, the dynamical models describing the $D$-meson decay amplitudes have to be further 
improved by introduction of unitarity, analyticity, chiral symmetry constraints and by inclusion 
of a hadronic input from studies of other reactions than the weak $D$-meson decays. 

\Acknowledgements
The author thanks Jean-Pierre Dedonder, Robert Kami\'nski and Benoit Loiseau for a long cooperation.
His participation in the CKM2014 Conference has been supported by the National Science Center 
(grant number UMO-2013/09/B/ST2/04382).

\end{document}

%% file: econfmacros.tex
\def\beq{\begin{equation}}
\def\eeq#1{\label{#1}\end{equation}}
\def\eeqn{\end{equation}}

\def\beqa{\begin{eqnarray}}
\def\eeqa#1{\label{#1}\end{eqnarray}}
\def\eeqan{\end{eqnarray}}

\let\bar=\overbar

\def\Dslash{\not{\hbox{\kern-4pt $D$}}}
\def\dslash{\not{\hbox{\kern-2pt $\del$}}}

\def\msb{{\bar{\ssstyle M \kern -1pt S}}}

